\documentclass{article}
\usepackage{psfig}
\usepackage{epsfig}
\date{}

\begin{document}

\title{\bf Casimir force between partially transparent plane mirrors}
\author{Ricardo M.\ Cavalcanti\footnote{E-mail: rmoritz@if.ufrj.br} \\
\small\it Instituto de F\'{\i}sica, Universidade Federal do
Rio de Janeiro \\
\small\it Caixa Postal 68528, 21945-970 Rio de Janeiro, RJ}
\maketitle

\begin{abstract}

We compute the Casimir energy of a real scalar field in the
presence of a pair of partially transparent plane mirrors, 
modeled by Dirac delta potentials.

\end{abstract}


\section{Introduction}

The Casimir effect \cite{casimir} consists in the modification of
the vacuum energy of a quantum field due to the presence of one or
more macroscopic bodies. Their influence on the field
is usually modeled by a boundary condition. In the most simple
example of the Casimir effect, one considers a field in the
presence of a pair of perfectly reflecting plane surfaces
(mirrors). In the case of a scalar field, they can be replaced by
Dirichlet boundary conditions, i.e, the field vanishes at the
reflecting surfaces.

Perfectly reflecting surfaces are an idealization; any real mirror
is transparent at high frequencies. It is natural, therefore, to
ask how the Casimir force between two real mirrors would be.
Jaekel e Reynaud \cite{jaekel} presented a general solution to the
problem of partially transparent plane mirrors, by introducing
reflection and transmission coefficients obeying conditions of
unitarity, causality, and transparency at high frequencies. Bordag
{\it et al}.\ \cite{bordag} investigated a specific model for the
mirrors, in which they are represented by Dirac delta potentials.
This model is revisited in the present work. Results are obtained
for generic $D$ spatial dimensions.


\section{Casimir energy}

Our starting point is the Lagrangian density
\begin{equation}
{\cal L}=\frac{1}{2}\,(\partial\phi)^2
+\frac{1}{2}\left[m^2+V(z)\right]\phi^2.
\end{equation}
We work in a $(D+1)$-dimensional Euclidean space-time with
coordinates $x=(\tau,{\bf x})$, where ${\bf x}=({\bf r},z)$ and
${\bf r}=(x_1,\ldots,x_{D-1})$; we also employ the natural system
of units, $\hbar=c=1$. The scalar potential $V(z)$ is composed of
a pair of one-dimensional $\delta$-functions with support at $z=0$
and $z=\ell$:
\begin{equation}
\label{V}
V(z)=\lambda\left[\delta(z)+\delta(z-\ell)\right].
\end{equation}
This potential models two partially transparent plane mirrors
separated by the distance $\ell$.

In order to compute the Casimir energy, we shall use the
following identity (see Appendix A):
\begin{equation}
\label{identity}
E=\lim_{\tau'\to \tau}\,\frac{\partial^2}
{\partial \tau^2}\,F(\tau,\tau'),
\end{equation}
where $F(\tau,\tau')$ is defined as
\begin{equation}
\label{F}
F(\tau,\tau')\equiv\int d^Dx\,G(\tau,{\bf x};\tau',{\bf x}).
\end{equation}
The Green's function $G$, by its turn, can be obtained
by solving the partial differential equation
\begin{equation}
\label{G}
\left[-\partial^2+m^2+V(z)\right]G(x,x')
=\delta^{(D+1)}(x-x').
\end{equation}
Fourier transforming Eq.\ (\ref{G}) in $\tau$ and {\bf r}
reduces it to an ordinary differential equation
for ${\cal G}$, the Fourier transform of $G$:
\begin{equation}
\label{Gred}
\left[-\partial_z^2+\omega^2+{\bf k}^2+m^2+V(z)\right]
{\cal G}(\omega,{\bf k};z,z')=\delta(z-z'),
\end{equation}
\begin{equation}
\label{FG}
G(x,x')=\int\frac{d\omega}{2\pi}\,e^{-i\omega(\tau-\tau')}
\int\frac{d^{D-1}k}{(2\pi)^{D-1}}\,
e^{i{\bf k}\cdot({\bf r}-{\bf r}')}\,
{\cal G}(\omega,{\bf k};z,z').
\end{equation}

As a preliminary step to solve Eq.\ (\ref{Gred}), let us
first solve it in the absence of the
external potential $V$, i.e.,
\begin{equation}
\left[-\partial_z^2+\omega^2+{\bf k}^2+m^2\right]
{\cal G}_0(\omega,{\bf k};z,z')=\delta(z-z').
\end{equation}
Using the method of Fourier transforms one obtains
\begin{equation}
{\cal G}_0(\omega,{\bf k};z,z')
=\int_{-\infty}^{\infty}\frac{dq}{2\pi}\,\frac{e^{iq(z-z')}}
{\omega^2+{\bf k}^2+q^2+m^2}
=\frac{e^{-\sigma|z-z'|}}{2\sigma}\,,
\end{equation}
where $\sigma\equiv\sqrt{\omega^2+{\bf k}^2+m^2}$.
Using the results of Appendix B, one then obtains the
following expression for ${\cal G}$ in terms of
${\cal G}_0$:
\begin{equation}
\label{G2}
{\cal G}(z,z')={\cal G}_0(z,z')-\sum_{j,k=1}^2
{\cal G}_0(z,z_j)\,({\cal M}^{-1})_{jk}\,
{\cal G}_0(z_k,z'),
\end{equation}
where $z_1=0$, $z_2=\ell$, and the matrix ${\cal M}$ is
given by
\begin{equation}
{\cal M}=\left(\begin{array}{cc}
\frac{1}{\lambda}+{\cal G}_0(0,0) & {\cal G}_0(0,\ell) \\
 & \\
{\cal G}_0(\ell,0) & \frac{1}{\lambda}+{\cal G}_0(\ell,\ell)
\end{array}\right)
=\left(\begin{array}{cc}
\frac{1}{\lambda}+\frac{1}{2\sigma} &
\frac{e^{-\sigma\ell}}{2\sigma} \\
 & \\
\frac{e^{-\sigma\ell}}{2\sigma} &
\frac{1}{\lambda}+\frac{1}{2\sigma}
\end{array}\right).
\end{equation}

Now we have all the ingredients to compute the Casimir energy. In
order to obtain a finite result, however, we must subtract (i) the
energy of the vacuum in the absence of the mirrors and (ii) the
self-energy of each mirror. These quantities are formally infinite,
but, since they do not depend on the distance between the mirrors, 
they do not contribute to the Casimir force. Thus the renormalized
Casimir energy is given by
\begin{eqnarray}
E_{\rm ren}\!\!\!\!&=&\!\!\!\!\lim_{\tau'\to\tau}\frac{\partial^2}
{\partial\tau^2}
\int d^Dx\int_{-\infty}^{\infty}\frac{d\omega}{2\pi}\,
e^{-i\omega(\tau-\tau')}
\int\frac{d^{D-1}k}{(2\pi)^{D-1}}\,{\cal G}_{\rm sub}(\omega,{\bf k};z,z)
\nonumber \\
&=&\!\!\!\!-L^{D-1}\int_{-\infty}^{\infty} dz
\int_{-\infty}^{\infty}\frac{d\omega}{2\pi}
\int\frac{d^{D-1}k}{(2\pi)^{D-1}}\,\omega^2\, {\cal
G}_{\rm sub}(\omega,{\bf k};z,z), \label{ER1}
\end{eqnarray}
where $L^{D-1}$ is the area of each plate and ${\cal G}_{\rm sub}=({\cal
G}-{\cal G}_0)-2\,({\cal G}_1-{\cal G}_0)$ (the first
subtraction removes the energy of the vacuum in the absence of the
mirrors, and the second one removes the self-energy of the mirrors;
${\cal G}_1$ denotes the Fourier transform of the Green's function
when only one mirror is present). The multiple integral in
(\ref{ER1}) can be reduced to a single one; the final result is
\begin{equation}
\label{ER2} 
E_{\rm ren}=-\frac{L^{D-1}}{(4\pi)^{D/2}
\Gamma\left(\frac{D}{2}+1\right)}\, \ell^{-D}\,
{\cal I}(m\ell,\lambda\ell),
\end{equation}
where
\begin{equation}
\label{I}
{\cal I}(a,b)\equiv\int_a^{\infty}ds\,\frac{(s^2-a^2)^{D/2}
\left(\frac{2s}{b}+\frac{2}{b}
+1\right)e^{-2s}}{\left(\frac{2s}{b}
+1\right)\left[\left(\frac{2s}{b}+1\right)^2-e^{-2s}\right]}.
\end{equation}
It agrees with Bordag {\it et al.}'s result for $D=3$ (the only
case they considered).

Let us examine in more detail the massless case ($m=0$), 
in which the Casimir force is long-ranged. In the limit 
$\lambda\to\infty$ one obtains
\begin{equation}
{\cal I}(0,\lambda\ell)\sim\frac{\Gamma(D+1)}{2^{D+1}}\,
\zeta(D+1)+O(1/\lambda\ell);
\end{equation}
combined with the pre-factor in (\ref{ER2}), this yields the
well-known expression \cite{casimir} for the Casimir energy of a
massless scalar field in the presence of two ideal (i.e., 
perfectly reflecting) plane mirrors in $D$ spatial dimensions:
\begin{equation}
\lim_{\lambda\to\infty}\,\frac{E_{\lambda}(\ell)}{L^{D-1}}
=-\frac{\Gamma(D+1)\,\zeta(D+1)}
{2^{2D+1}\pi^{D/2}\Gamma\left(\frac{D}{2}+1\right)}\,
\ell^{-D}.
\end{equation}

Figs.\ 1 and 2 depict the ratio $R\equiv
E_{\lambda}(\ell)/E_{\infty}(\ell)$ as a function of the parameter
$x\equiv\lambda\ell$ in the massless case for $D=1,2,3$. One notices that
$E_{\lambda}(\ell)$ is always smaller, in absolute value, than
$E_{\infty}(\ell)$; in particular, the ratio of the former to the
latter approaches zero as $\lambda\ell\to 0$.


\begin{figure}
\begin{center}
\psfig{figure=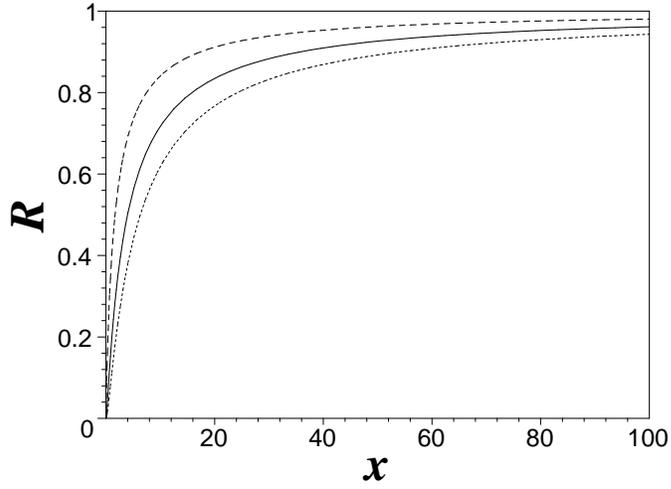,width=9cm,angle=0}
\vspace{1.0cm}
\caption{$R\equiv E_{\lambda}(\ell)/E_{\infty}(\ell)$
{\it vs}.\ $x\equiv\lambda\ell$ in the massless case
for $D=1$ (long-dashed line),
$D=2$ (solid line), and $D=3$ (short-dashed line).}
\end{center}
\end{figure}

\begin{figure}
\begin{center}
\psfig{figure=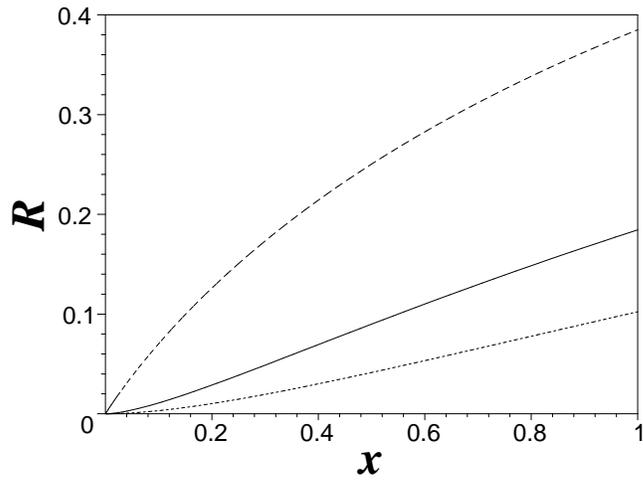,width=9cm,angle=0}
\vspace{1.0cm}
\caption{Zoom of Fig.\ 1 near the origin.}
\end{center}
\end{figure}


\section*{Acknowledgments}

This work grew out of discussions with Luiz C.\ de Albuquerque,
whom I gratefully acknowledge. I also acknowledge the hospitality
of the Departamento de F\'{\i}sica Matem\'atica, Universidade de
S\~ao Paulo, where this work was initiated, and the financial
support from FAPESP and FAPERJ. Last, but no least, I thank
C.\ Farina and M.\ V.\ Cougo-Pinto for a critical reading of
the manuscript.


\renewcommand{\theequation}{\Alph{section}\arabic{equation}}

\section*{Appendix A}
\setcounter{equation}{0}
\setcounter{section}{1}

Here I prove the identity (\ref{identity}).
The formal solution to Eq.\ (\ref{G}) reads
\begin{equation}
\label{G'}
G(\tau,{\bf x};\tau',{\bf x}')=\int\frac{d\omega}{2\pi}\,
e^{-i\omega(\tau-\tau')}\sum_n\frac{\varphi_n({\bf x})\,
\varphi_n^*({\bf x}')}{\omega^2+\omega_n^2},
\end{equation}
where $\varphi_n({\bf x})$ and $\omega_n$ satisfy
\begin{equation}
(-\nabla^2+m^2+V)\,\varphi_n({\bf x})
=\omega_n^2\,\varphi_n({\bf x}),
\end{equation}
\begin{equation}
\label{norm}
\int d^Dx\,\varphi_n^*({\bf x})\,\varphi_n({\bf x})=1.
\end{equation}
Inserting (\ref{G'}) into (\ref{F}), and using
the normalization condition (\ref{norm}), one obtains
\begin{equation}
F(\tau,\tau')=\sum_n\int\frac{d\omega}{2\pi}\,
\frac{e^{-i\omega(\tau-\tau')}}{\omega^2+\omega_n^2}
=\sum_n\frac{e^{-\omega_n|\tau-\tau'|}}{2\,\omega_n}
\qquad(\omega_n>0).
\end{equation}
Therefore,
\begin{equation}
\lim_{\tau'\to \tau}\,\frac{\partial^2}{\partial \tau^2}\,
F(\tau,\tau')
=\lim_{\tau'\to \tau}\,\sum_n\frac{1}{2}\,\omega_n\,
e^{-\omega_n|\tau-\tau'|}=E,
\end{equation}
as promised.


\section*{Appendix B}
\setcounter{equation}{0}
\setcounter{section}{2}

The differential equation
\begin{equation}
\label{g}
\left[-\partial_x^2+\sigma^2+V(x)\right]
{\cal G}(x,x')=\delta(x-x')
\end{equation}
can be reexpressed as an integral equation --- the
Lippmann-Schwinger equation:
\begin{equation}
\label{LS1}
{\cal G}(x,x')={\cal G}_0(x,x')-\int dy\,{\cal G}_0(x,y)\,
V(y)\,{\cal G}(y,x'),
\end{equation}
where ${\cal G}_0(x,x')$ satisfies
\begin{equation}
\left[-\partial_x^2+\sigma^2\right]
{\cal G}_0(x,x')=\delta(x-x').
\end{equation}
If $V(x)=\lambda\,\delta(x-x_0)$, the integral in Eq.\ (\ref{LS1})
can be performed trivially, yielding
\begin{equation}
\label{LS2}
{\cal G}(x,x')={\cal G}_0(x,x')-\lambda\,{\cal G}_0(x,x_0)\,
{\cal G}(x_0,x'),
\end{equation}
Setting $x=x_0$ in Eq.\ (\ref{LS2}), solving for ${\cal
G}(x_0,x')$, and inserting the result back into (\ref{LS2}) one
finally obtains \cite{hennig}
\begin{equation}
{\cal G}(x,x')={\cal G}_0(x,x')-\frac{{\cal G}_0(x,x_0)\,
{\cal G}_0(x_0,x')}{\frac{1}{\lambda}+{\cal G}_0(x_0,x_0)}.
\end{equation}

This result can be easily generalized to a potential composed of
$N$ $\delta$'s: if
\begin{equation}
V(x)=\sum_{j=1}^{N}\lambda_j\,\delta(x-x_j),
\end{equation}
the solution to Eq.\ (\ref{g}) is given by
\begin{equation}
\label{GN}
{\cal G}(x,x')={\cal G}_0(x,x')-\sum_{j,k=1}^N {\cal
G}_0(x,x_j)\,({\cal M}^{-1})_{jk}\,{\cal G}_0(x_k,x'),
\end{equation}
where ${\cal M}$ is the $N \times N$ matrix whose elements are
given by
\begin{equation}
\label{M}
{\cal M}_{jk}=\frac{1}{\lambda_j}\,\delta_{jk} +{\cal
G}_0(x_j,x_k).
\end{equation}



\begin{thebibliography}{99}

\bibitem{casimir}
G.\ Plunien, B.\ Muller, and W.\ Greiner,
Phys.\ Rep.\ {\bf 134}, 87 (1986);
V.\ M.\ Mostepanenko and N.\ N.\ Trunov,
Sov.\ Phys.\ Usp.\ {\bf 31}, 965 (1988);
M.\ Bordag, U.\ Mohideen, and V.\ M.\ Mostepanenko,
Phys.\ Rep.\ {\bf 353}, 1 (2001) [quant-ph/0106045].

\bibitem{jaekel} M.\ T.\ Jaekel and S.\ Reynaud,
Journal de Physique I {\bf 1}, 1395 (1991) [quant-ph/0101067].

\bibitem{bordag} M.\ Bordag, D.\ Hennig, and D.\ Robaschik,
J.\ Phys.\ A: Math.\ Gen.\ {\bf 25}, 4483 (1992).

\bibitem{hennig} Alternative derivations of this result
can be found in 
S.\ Albeverio, F.\ Gesztesy, R.\ H{\o}egh-Krohn, and H.\ Holden,
{\it Solvable Models in Quantum Mechanics} (Springer-Verlag,
New York, 1988);
C.\ Grosche, J.\ Phys.\ A: Math.\ Gen.\ {\bf 23}, 5205 (1990);
Annalen Phys.\ {\bf 2}, 557 (1993) [hep-th/9302055]; 
D.\ Hennig and D.\ Robaschik, 
Phys.\ Lett.\ {\bf 151A}, 209 (1990).

\end{thebibliography}
\end{document}